\documentclass[aps,pre,groupedaddress,
notitlepage,11pt]{revtex4-2}
\usepackage[utf8]{inputenc}
\usepackage{subfig}
\RequirePackage{fix-cm}
\usepackage[table]{xcolor}
\usepackage{graphicx}
\usepackage{braket}
\usepackage{float}
\usepackage{ragged2e}
\usepackage{array}
\usepackage{dcolumn}
\usepackage[colorlinks=true, citecolor=blue]{hyperref}
\usepackage{hhline}
\usepackage{amssymb}
\usepackage{amsmath}
\usepackage[utf8]{inputenc}
\usepackage{tikz}
\usepackage[T1]{fontenc}
\graphicspath{{dropbox/}}
\def\beq{\begin{equation}}
	\def\eeq{\end{equation}}
\begin{document}
	\large
\title{Virtual temperatures as a key
quantifier for passive states in
quantum thermodynamic processes}
    \author{Sachin Sonkar} 
	\email[e-mail: ]{ph19077@iisermohali.ac.in}
	\affiliation{ Department of Physical Sciences, 
		Indian Institute of Science Education and Research Mohali,  
		Sector 81, S.A.S. Nagar, Manauli PO 140306, Punjab, India} 
	\author{Ramandeep S. Johal} 
	\email[e-mail: ]{rsjohal@iisermohali.ac.in}
	\affiliation{ Department of Physical Sciences, 
		Indian Institute of Science Education and Research Mohali,  
		Sector 81, S.A.S. Nagar, Manauli PO 140306, Punjab, India}
\begin{abstract}
We analyze the role of virtual temperatures for passive quantum states 
through the lens of majorization theory.
A mean temperature over 
the virtual temperatures of adjacent energy levels is defined 
to compare the passive states of the system resulting from isoenergetic and isoentropic transformations.
The role of the minimum 
and the maximum (min-max) 
values of the virtual temperatures in determining the direction of heat flow between the 
system and the environment is argued based on majorization relations. 
We characterize the intermediate passive states in a quantum Otto engine using these virtual temperatures and derive an upper bound for the Otto efficiency that can be expressed in terms of the min-max virtual temperatures of the working medium. An explicit example of the coupled-spins system
is worked out.
Moreover, virtual temperatures serve to draw interesting parallels between the quantum thermodynamic processes and their classical counterparts. Thus, virtual temperature emerges as a key operational quantity linking passivity and majorization to the optimal performance of quantum thermal machines.
\end{abstract}
 \maketitle
\section{Introduction}
Heat exchange is a fundamental process in natural phenomena,  
with wide-ranging implications in basic and applied sciences as well as for human experience. It finds its
proper formulation within the purview of thermodynamics 
whereby a spontaneous flow of heat can be understood as the flow of (disordered) energy
from higher to lower temperatures.  
On the other hand, the notion of temperature is 
well-understood primarily for macroscopic systems
in equilibrium. 
The notion usually becomes ill-defined and elusive
for systems away from equilibrium \cite{CasasVazquez2003}, finite-sized systems 
\cite{Hilbert2014,Campisi2015} or interacting systems 
\cite{HernandezSantana2015,Kliesch2014,Hartmann2004}. 
 When extending the discussion to non-equilibrium quantum systems, it may be useful to introduce the notion of an effective temperature as a means to describe and quantify thermodynamic properties 
 ~\cite{Mahlerbook2008, Johal2009, Puglisi2017,Hsiang2021,Allahverdyan2023,Sobolev2018,Ghonge2018}.
A key question of broad significance is: How does a non-equilibrium quantum system behave when put in 
thermal contact with an environment at a given 
temperature? We focus on this question for a special 
class of states called passsive states which 
are specified by a set of positive-valued, 
virtual temperatures \cite{Skrzypczyk2015}. We define a mean virtual temperature 
and show its utility to compare passivity 
of states obtained under isoenergetic and isoentropic
processes. 

The concept of passivity was first introduced in a seminal work \cite{Pusz1978}  for identifying equilibrium states in general quantum systems. A state is defined to be  passive if no work can be extracted from a thermally isolated system through any cyclic unitary process. Within this framework, it was proved that both $\beta$–KMS states and ground states are completely passive, meaning that even an infinite number of copies cannot be exploited to extract work.
Building on this foundation, Ref. \cite{Skrzypczyk2015} further clarified the distinction between passive and completely passive states using the concept of virtual temperatures \cite{Brunner2012},  
which were further studied in Refs. \cite{Silva2016,Mitchison2019} along with the notion of a virtual qubit.  The energy instability of a passive state was analyzed \cite{Sparaciari2017}, and its characterization from a geometric perspective has been explored \cite{Nikolaos2021}.
More recently, an operational definition of temperature was proposed for non-equilibrium quantum systems \cite{Brunner2023}, extending beyond the standard energy-matching temperature.

In classical thermodynamics, two kinds of
processes occupy a fundamental significance.  
Whereas the entropy-conserving processes are reversible, the energy-conserving processes usually lead to entropy generation. 
A classical macroscopic system, prepared 
in a non-equilibrium state, 
can be brought to a final equilibrium state at a specific temperature 
by either of the above processes. It is known 
that the entropy-conserving final temperature
is, in general, lower than the energy-conserving 
final temperature \cite{JohalAJP2023}. 
The corresponding quantum thermodynamic processes
are respectively modelled by cyclic unitaries and 
unital CPTP operations  \cite{NielsenBook, Sagawa2020}. 
While in the quantum case, the transformed states 
may not attain a well-defined temperature, 
we can define a mean virtual temperature 
for a passive state, which is the weighted
mean of the virtual temperatures 
between all the adjacent energy levels.
We show that this mean temperature plays an  
equivalent role of the final temperature
in classical thermodynamics, 
after state transformations under isoenergetic and isoentropic processes.  
Further, we give a thermodynamic interpretation 
for the criterion 
that a passive state will always lose energy
to a heat reservoir at a temperature equal 
to or less than the minimum virtual temperature
of the state \cite{Brunner2023}. Alternately, the state will 
gain energy in thermalizing with an environment
at or greater than then maximum virtual 
temperature of the state. 
Finally, we use this criterion in  
a quantum Otto cycle to characterize
the passive state resulting from  
a quantum adiabatic process. 
We show that the min-max virtual temperatures of these
passive states determine an upper bound for the 
efficiency of the Otto cycle, which is explicitly
calculated for a two-qubit interacting working medium.

Mathematically, the above findings of the paper  
are related by a common thread, that is the theory of majorization. 
A probability distribution $P = (p_1,..., p_n)$
is said to be majorized by the distribution 
$Q = (q_1,..., q_n)$, the distributions 
being ordered in non-increasing sense, 
if the following set of relations are satisfied:
$M_i = \sum_{j=i+1}^{n} (p_j - q_j) \ge 0$, where 
$i=1,..., n-1$. Intuitively, it implies that the distribution 
$P$ is more spread out than $Q$.
In particular, these relations imply $p_1 \le q_1$ and 
$p_n \ge q_n$. 
The majorization relation described above is denoted as $P \prec Q$.
The concept of majorization and its generalizations constitute  an important analytical tool driving the current theoretical developments \cite{Marshall2011, Bhatia1996, Sagawa2020, Joe1990,Ruch1980,Torun2023}.  In quantum information, transformations between pure bipartite states under local operations and classical communication are completely characterized by majorization of their Schmidt coefficients  \cite{Nielsen1999, Nielsen2001}. More generally, majorization determines the feasibility of state transformations transformation in the various resource theories \cite{Horodecki2013,Gour2018}. Further, majorization has found various applications in quantum thermodynamic tasks \cite{Uttam2021,Rethinasamy2020,Lostaglio2022,Alimuddin2019,ADOJunior2022,Sachin2023}.

\section{Min-max virtual temperatures}
\label{vt}
Consider an $n$-level quantum system, with 
Hamiltonian $H = \sum_{k=1}^{n}E_k |k\rangle \langle k|$, 
in a quantum state described by the density matrix 
$\rho^p = \sum_{k=1}^{n}p_k |k\rangle \langle k|$.  
Clearly, we have $[\rho^p,H]=0$. Let us assume that
the energies and their  occupation probabilities are ordered as 
 $E_k < E_{k+1}$ with $p_k \ge p_{k+1}$.
Such a state is said to be passive since 
its mean energy, $U(\rho^p) = \sum_k E_k p_k$, cannot be lowered by applying some cyclic unitary. 
In other words, no work can be extracted from 
the system in this manner.
The completely passive states 
are the thermal states: $\rho(T) \sim e^{-H/k_{\rm B}T}$ ($k_{\rm B}^{}=1$), which are parametrized by a single temperature $T>0$. 
A passive state may be characterized by a set of so-called 
virtual temperatures \cite{Brunner2012}. 
The virtual temperature corresponding
to a pair of energy levels $(i,j)$ is defined as
$T_{ij} = (E_j - E_i)/\ln (p_i/p_j) >0$. 
For our purpose, we also consider 
the subset of virtual temperatures corresponding to  
 pairs of adjacent levels $(i,i+1)$, defined as 
\begin{equation}
   T_i =  \frac{\omega_i}{\ln p_i -\ln p_{i+1} }, 
   \quad i=1,...,n-1
   \label{vti}
\end{equation}
with the energy gap $\omega_i = E_{i+1} - E_i$. 
Clearly, all $T_i$
coincide in the case of a thermal state $p_i \sim e^{-E_i/T}$. 
We first establish the following result.
\par\noindent
{\it Lemma}: The min-max virtual temperatures for a passive state, 
$T_{\rm min} = \underset{i\neq j}{\rm min}\; T_{ij}$ and  $T_{\rm max} = \underset{i\neq j}{\rm max}\; T_{ij}$,    
    are contained in the subset $\{T_{i}\}$, i.e. they 
belong to some adjacent pairs of levels.

To prove the above, it is sufficient to note that the virtual 
temperature between a non-adjacent pair of levels
cannot have an extremal value. 
Consider $T_{ij}$ between any pair of non-adjacent levels 
($j<i$) separated by  the gap $\omega_{ij}=E_j-E_i = \sum_{k=i}^{j-1} \omega_k$, and with  
$\{T_k|k=i,...,j-1\}$.
Then, we can write 
${T}_{ij}=\left(\sum_{k=i}^{j-1} {\Omega_k}/{T_k}\right)^{-1}$,   
where the weights, $\Omega_k={\omega_k}/{\sum_{k=i}^{j-1} \omega_k} >0$, satisfy  $\sum_k \Omega_k=1$.
In other words, $T_{ij}$ is the weighted harmonic mean of the virtual temperatures of the adjacent levels contained only within the gap $\omega_{ij}$. Therefore,
by the property of a mean, ${T}_{ij}$ is strictly bounded 
by the minimum and the maximum values of the $\{T_k\}$, 
implying that it cannot be an extremal value. 
It follows that in order to determine the min-max virtual 
temperatures in the case of a multi-level system,
we need to consider only the subset $\{T_i\}_{i=1,...,n-1}$.

\section{Mean virtual temperature}
\label{mvt}
The above result suggests a natural definition of the mean virtual temperature for a  passive state, given as the weighted 
harmonic mean of the virtual temperatures 
for {\it all} adjacent pairs of levels, i.e. $\{T_k|k=1,...,n-1\}$:
\begin{equation}
\widetilde{T}=\left(\sum_{k=1}^{n-1} \frac{\Omega_k}{T_k}\right)^{-1},    \label{t1n}
\end{equation}
where $\Omega_k={\omega_k}/{\sum_{k=1}^{n-1}\omega_k}$.
Therefore, $\widetilde{T}$ is bounded as $T_{\rm max}\geq \widetilde{T} \geq T_{\rm min}$. Further,
it  is  equal to the virtual temperature  between the ground and the top levels, i.e. $T_{1n}$. 

\begin{figure}[h!]
     \centering
     \includegraphics[width=0.8\linewidth]{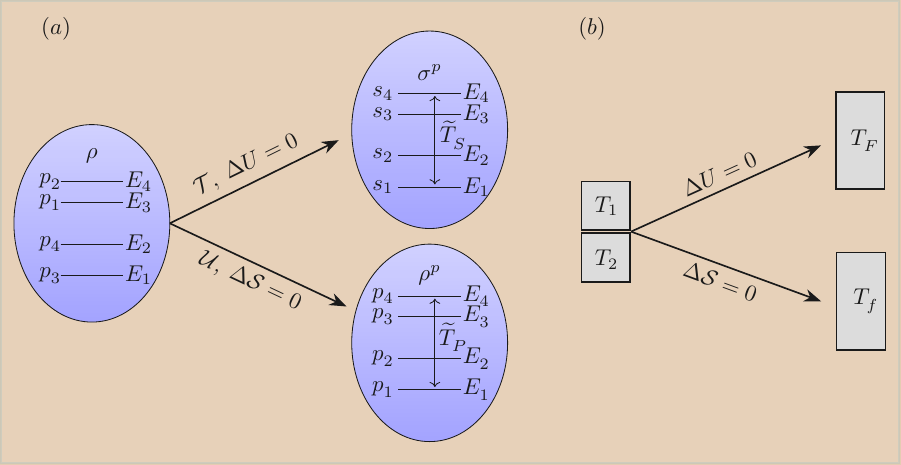}
     \caption{(a) An active state $\rho$, with Hamiltonian $H$,   may be 
     transformed into a passive state $\sigma^p$  via an isoenergetic process ($\Delta U=0$) using a 
     unital CPTP map $\mathcal{T}(\rho) = \sigma^p$, 
     or into a passive state $\rho^p$ via a cyclic
     unitary, 
     $\mathcal{U} \rho\; \mathcal{U^{\dagger}} = \rho^p$, 
     which preserves the von Neumann entropy 
     ($\Delta \mathcal{S} = 0$).
     For simplicity, we take the initial active state 
     with zero coherence.
     The majorization relation $\rho^p \succ \sigma^p$
     implies that the corresponding mean virtual temperatures satisfy: 
     $\widetilde{T}_S > \widetilde{T}_P$. (b) The classical analogue where an initial nonequilibrium state,  modelled as two subsystems at unequal temperatures
     $T_1$ and $T_2$, reaches the final equilibrium
     state at a specific temperature, via an 
     isoenergetic process (final temperature $T_F$)
     or an isoentropic process (final temperature $T_f$),
     satisfying  $T_F > T_f$.}
\label{Process}
 \end{figure}

Next, we observe that given two 
passive states of a quantum system, $\rho^p$ and $\sigma^p$,  
 with the respective energy level distributions $P$ and $S$
satisfying the relation $S\prec P$, 
 the mean virtual temperatures
of the two states satisfy 
$\widetilde{T}_P < \widetilde{T}_S$. To see this, 
suppose that the system is initially in a non-passive or active
state $\rho$. For simplicity, we consider $\rho$ to be 
diagonal in the energy basis. 
Let it be brought to the passive state 
$\rho^p = \mathcal{U} \rho \mathcal{U}^{\dagger}$, 
by applying a cyclic unitary $\mathcal{U}$ to the system Hamiltonian i.e. $H = \mathcal{U} H \mathcal{U}^{\dagger}$.
Since the eigenvalues of the state are preserved, this
transformation involves only the rearrangement 
of the probabilities such that the final probabilities 
are ordered in a non-increasing sense. 
Such an evolution maximally lowers the energy of
the system 
while conserving its von Neumann entropy, 
$\mathcal{S}(\rho) = {\rm Tr} (\rho \ln \rho)$ \cite{Allahverdyan2004}.  
On the other hand, we may subject the state $\rho$ 
to an energy-preserving or isoenergetic operation, 
 which is a unital CPTP map 
transforming $\rho$ into a state $\sigma^p$ at the same initial mean energy:
$ {\rm Tr}[\sigma^p H] = {\rm Tr}[\rho H]$. Such a
state transformation implies the majorization relation:
 $ \rho \succ \sigma^p$ \cite{Ali2020,Chiribella2017}. 
Further, we require that the transformed
 state $\sigma^p$ be passive, so that the 
 the final state is diagonal in the energy basis and
 the occupation probabilities of $\sigma^p$ are also ordered
in a non-increasing sense: $S^{\downarrow} = \{ s_1,...,s_n \}$.   
Due to the majorization inequalities, we have:
$s_n > p_n$ and $s_1< p_1$,  or  $s_1/s_n < p_1/p_n$,
which can be written as  
$e^{\omega_{1n}/\widetilde{T}_S} < e^{\omega_{1n}/\widetilde{T}_P}$, which  implies  
$\widetilde{T}_S> \widetilde{T}_P$.
Thus, we find that the mean virtual temperature $\widetilde{T}_P$
of the passive state after ergotropy extraction 
(an isoentropic process) 
is lower than the mean virtual temperature $\widetilde{T}_S$ of 
the passive state obtained after an 
isoenergetic process (see Fig. \ref{Process}a) . 

The above statement may be 
regarded as the quantum thermodynamic analog
of the classical situation. For this, we consider a classical, non-equilibrium 
system which may be composed of two 
equilibrium subsystems 
at temperatures $T_1$ and $T_2$. The 
composite system can be brought to equilibrium 
at a certain temperature $T_f$ by coupling
the two subsystems via a perfect engine
that extracts the maximum work while conserving
the total (thermodynamic) entropy. We call $T_f$ as the entropy-conserving 
temperature (see Fig. \ref{Process}b). On the other hand, we can 
visualize an energy-conserving process
by which the two subsystems are put in 
mutual thermal contact so that heat may flow
from the hotter to the colder subsystem, till
they attain a common temperature $T_F$,
called the energy-conserving temperature.
Now, it can be shown quite generally 
that $T_F > T_f$ \cite{JohalAJP2023, Narang2023}, 
based on the assumption of
positive heat capacities for the subsystems.

Comparing with the situation with 
$\rho^p$ and $\sigma^p$ above, the classical, 
reversible work extraction corresponds
to the extraction of ergotropy via a unitary
process, while the energy exchange via 
a thermally insulated process---in which 
the final entropy increases, corresponds
to the application of a unital CPTP map.
It is important to note that after the 
application of the unital map, the system may still
be left in an active state. The present
analogy with classical thermodynamics is applicable 
if the final state is a passive state,
for only then  all the virtual temperatures,
and hence, the mean virtual 
temperature is positive, making it possible to compare 
$\widetilde{T}_S$ and $\widetilde{T}_P$.
The requirement of a final passive state 
implies that the intial active state $\rho$
dissipates or loses its potential for work,
which, in classical terms, corresponds to 
an initial non-equilibrium state attaining
a final, uniform temperature. 
Unlike the classical systems, the quantum 
systems, in general, do not attain equilibrium states,
either after ergotropy extraction, or after
passing through the unital channel. 
In this regard, the mean virtual temperature  
serves as an indicator  
equivalent to the final equilibrium 
temperature in the classical, macroscopic case,  
with an inequality $\widetilde{T}_S> \widetilde{T}_P$ 
analogous to the 
classical inequality $T_F > T_f$. 

\section{The direction of heat flow}
\label{heatflow}
In macroscopic thermodynamics, heat flows in the direction from hot to cold i.e. if the
temperature is well defined locally, then a gradient
of temperatures decides the direction
of the flow of heat. 
The question is: Given a quantum system in a passive state $\rho^p$,
what determines the heat flow when the system
is put in contact with a thermal environment at 
 temperature $T$? It was shown in Ref.~\cite{Brunner2023}
that for the heat to flow from the system toward the environment, $T_{\rm min} \geq T$ provides a 
sufficient criterion. Similarly, the direction 
of heat flow will always be from the environment
to the system, if $T \geq T_{\rm max}$. 
The proof there was based 
on resource-theoretic arguments. In the following, we  relook at this result through the lens of majorization.

Let $P$ denote 
the occupation distribution for the energy levels $E_i$
of the system with the Hamiltonian $H$, in the
given passive state $\rho^p$, while 
$Q\equiv Q(T_{\min})$ denotes the canonical 
distribution of the system in thermal state
$\rho(T_{\min})$, where $T_{\min}$ is 
the minimum virtual temperature in the state $\rho^p$.
We show in appendix A that $P \prec Q(T_{\min})$,
implying that there exists a randomizing map 
which transforms $\rho^p$ into $\rho(T_{\min})$.
A simple example of such a transformation
is the thermal contact between the system
and an environment at temperature $T_{\min}$. Then,
the task is to determine the direction of average heat
flow during the consequent thermalization of the system. 

Intuitively, we note that the majorization
implies $Q(T_{\min})$
is less spread out than $P$, and so 
the entropies satisfy: $\mathcal{S}(\rho(T_{\min})) < \mathcal{S}(\rho^p)$.
In other words, upon thermal contact with environment
at $T_{\min}$, the final entropy of the system is less
than the initial, or, heat should 
flow out of the system and so $U(\rho^p) \ge U(\rho(T_{\min}))$. 
A more formal proof is given in Appendix \ref{BoundME}. 
Along similar lines, we can prove that
$U(\rho(T_{\rm max}^{})) \geq U(\rho^p)$. Combining these 
two inequalities, we conclude that the min-max virtual temperatures can be used to bound the mean energy $U(\rho^p)$, as
\begin{equation}
  U(\rho(T_{\rm max}^{})) \geq U(\rho^p) \geq  
  U(\rho(T_{\rm min}^{})).
  \label{U3ineq}
\end{equation}
Then, for any environment temperature
  $T$ lower than $T_{\rm min}$, we have 
  $U(\rho^p) \geq  
  U(\rho(T_{\rm min}^{})) > U(\rho(T))$. So,  
  the system must lose energy equal to  
  $U(\rho^p)-U(\rho(T)) >0$, in the form of heat to
  the environment.  A similar argument holds for 
  $T \geq T_{\rm max}$, where the direction of 
  the heat flow will be reversed. 
  
  On the other hand, 
  for intermediate values $T_{\rm max} \geq T \geq T_{\rm min}$,
  the direction of the heat flow is not determined by the 
  above arguments. A specific protocol may be devised by resonantly coupling
  a pair of levels of the system with a pair of levels
  in the environment via an energy-conserving unitary process,
  which directs the flow of heat \cite{Brunner2023}.  
Usually, an effective temperature $T^*$ for the passive  state is
defined as the temperature of the Gibbs state 
 $\rho (T^*)$ that yields the same mean energy: 
 $U(\rho (T^*)) = U(\rho^p)$. 
Considering that the heat capacity for a thermal state
is positive,
$\partial U(\rho(T))/\partial T > 0$,  
we conclude from Eq. (\ref{U3ineq}) that 
$T_{\rm max}^{} \geq T^* \geq T_{\rm min}^{}$.
The temperature $T^*$ plays a useful role 
in the discussion of thermalization in
isolated quantum-many body systems. 
The effective temperature is also 
connected with the min-max 
virtual temperatures in the asymptotic limit \cite{Brunner2023}.
In the present context, 
 if we know the effective temperature $T^*$
  for the state $\rho^p$, 
  then we can use it to decide the direction of 
  heat flow for the case $T_{\rm max} \geq T \geq T_{\rm min}$,  where $T$ is the temperature of the 
  environment. Thus, if 
  $T^* > T$ ($T^* < T$), then we can still say that
  heat will always flow from (towards) the system.

\section{Quantum Otto efficiency}
\label{Ottoeff}
\begin{figure}[h!]
     \centering     \includegraphics[width=0.9\linewidth]{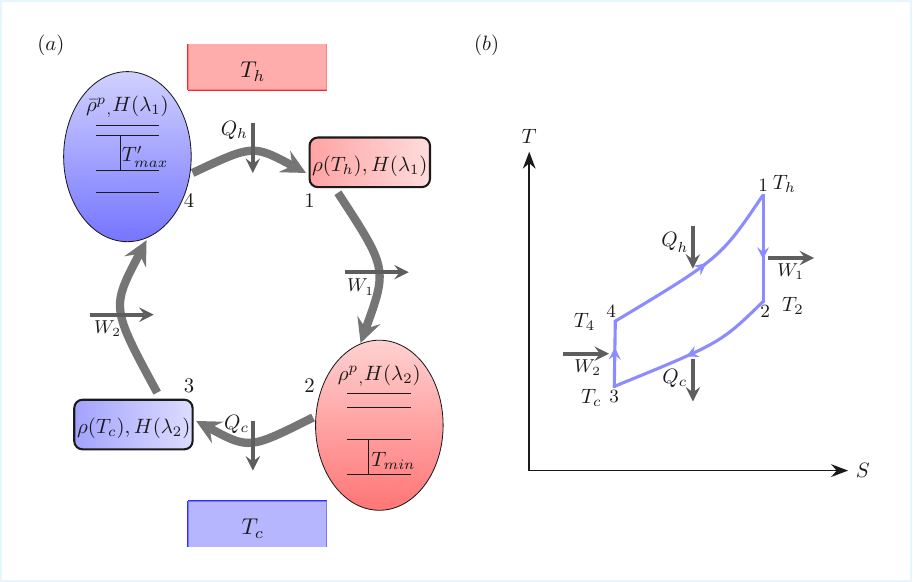}
    \caption{
(a) Schematic of a quantum Otto engine.
A quantum working medium with Hamiltonian $H(\lambda_1)$ is prepared
in the thermal state $\rho(T_h)$ with a hot reservoir. 
The first quantum adiabatic process ($1 \rightarrow 2$)
transforms the system into a passive state $\rho^p$ 
with Hamiltonian $H(\lambda_2)$. The system thermalizes with
the cold reservoir attaining the state $\rho(T_c)$.
The second quantum adiabatic process ($3 \rightarrow 4$)
transforms the system into the passive state $\bar\rho^p$ 
with Hamiltonian $H(\lambda_1)$, finally attaining the state $\rho(T_h)$.
$T_{\min}$ and $T_{\max}'$ are the minimum and maximum virtual temperatures
associated with $\rho^p$ and $\bar{\rho}^p$, respectively,
which can be used to fix the direction of heat flows $Q_{h,c}$ between
the system and the reservoir, as well as to upper bound 
the efficiency of the cycle. 
(b) shows the classical Otto cycle on a $TS-$plane, 
with a well-defined temperature 
$T_2$ ($T_4$) after the first (second) adiabatic stroke,
respectively.
}
 \label{QOC}
 \end{figure}

Next, we consider a quantum Otto engine 
to highlight the role of virtual temperatures in determining the upper bound of the Otto efficiency. A quantum Otto cycle, operating between two heat reservoirs at temperatures $T_h>T_c$, consists of a quantum working medium undergoing
four strokes, alternating between adiabatic and isochoric processes \cite{Quan2007}.
We first summarize the various steps of the heat cycle (see also Fig. \ref{QOC}a).
Consider the working medium as an $n$-level quantum system with Hamiltonian $
H(\lambda_1) = \sum_k E_k |\psi_k\rangle\langle \psi_k|$. Initially, the system starts in a thermal state with the hot reservoir
\begin{equation}
\rho_h^{} = \frac{e^{-H(\lambda_1)/T_h}}{{\rm Tr}(e^{-H(\lambda_1)/T_h})},
\qquad 
p_k = \frac{e^{-E_k/T_h}}{\sum_k e^{-E_k/T_h}}.
\label{canpk}
\end{equation}
In the first adiabatic stroke, the system is isolated from the hot reservoir and the Hamiltonian is changed slowly from $H(\lambda_1)
\to H(\lambda_2)$,
whereby the energy levels change from $E_k$ to $E_k'$.
Due to the quantum adiabatic theorem~\cite{Fock1928}, no transitions occur between instantaneous energy eigenstates, keeping the occupation probabilities $p_k$ unchanged. 
Further, we assume no-level crossing in this transformation. 
In the next stroke, keeping $E_k'$ fixed,
the system thermalizes with the cold reservoir to
reach the state 
    \begin{equation}
    \rho_c^{} = \frac{e^{-H(\lambda_2)/T_c}}{{\rm Tr}(e^{-H(\lambda_2)/T_c})}, \qquad p_k' = \frac{e^{-E_k'/T_c}}{\sum_k e^{-E_k'/T_c}}.
     \end{equation}
The system is again isolated for the second adiabatic stroke and the Hamiltonian is slowly restored, $H(\lambda_2) \to H(\lambda_1)$, implying $E_k' \to E_k$, with populations $p_k'$ remaining unchanged.
 Finally, the system is brought in contact with the hot reservoir, restoring the initial thermal state $\rho_h^{}$. 

The heat exchanged with the hot and the cold  reservoir during an isochoric process
is given by the energy difference between the final and initial states during the process:
$Q_h = \sum_{k=1}^n E_k (p_k - p_k')$, and  
$Q_c = -\sum_{k=1}^n E_k' (p_k - p_k')$.
It is convenient to rewrite these expressions as \cite{Sachin2025}
\begin{equation}
Q_h=\sum_{i=1}^{n-1}\omega_i M_i,~Q_c=-\sum_{i=1}^{n-1}\omega_i'M_i,
\label{qhc}
\end{equation}
where $\omega_i (\omega_i')$ are the energy gaps at
the start (end) of the first adiabatic stroke,
and $M_i=\sum_{j=i+1}^{n-1}(p_j-p_j')$, a quantity
analogous to $\chi_m$ in appendix A. 
Now, the operation as an engine requires $Q_h > 0$ and $Q_c <0$.
To ensure the correct direction of heat flow
when the system thermalizes with a reservoir, 
 we consider the system state after the first adiabatic stroke, 
$\rho = \sum_k p_k \, |\psi_k'\rangle \langle \psi_k'|$,
where $p_k$ are given by Eq. (\ref{canpk})
and $|\psi_k'\rangle$  denote the eigenstates
of the Hamiltonian $H(\lambda_2)$, with eigenvalues $E_k'$.
In general, $p_k$ cannot be expressed in the form ${e^{-E_k'/T}}/{\sum_k e^{-E_k'/T}}$, for some temperature  $T$.
In other words, $\rho$ represents a non-equilibrium, passive state, since it commutes with the Hamiltonian $H(\lambda_2)$, and its populations $p_k$ are arranged, as in the initial
state, in the descending order
($p_k > p_{k+1}$ for $E_k' < E_{k+1}'$), due to no level-crossing.

Now, for the passive state obtained after the first adiabatic stroke, the virtual temperature $T_i$ between the pair of levels ($i,i+1$) is given by
$T_i = \omega_i'/ \ln (p_i/p_{i+1})=T_h(\omega_i'/\omega_i)$,
where Eq. (\ref{canpk}) is used.
The minimum virtual temperature 
of the state $\rho$
can then be given as 
\begin{equation}
T_{\rm min} = T_h. \underset{i}{\rm min} \left(
\frac{\omega_i'}{\omega_i}\right).
\label{tmin}
\end{equation}
From the arguments in the 
previous section and considering $T_{\rm min} > T_c$,  
we can easily infer that heat is released by the system
to the cold reservoir, or $Q_c <0$. Alternately, in terms of majorization, we can say that a thermal distribution 
at a colder temperature majorizes the thermal distribution
at a hotter temperature, yielding $Q(T_{\min}) \prec P'(T_c) 
\equiv P'$. 
We have already seen above that $P \prec Q(T_{\min})$.
Therefore, by the transitive property, 
we infer that $P \prec P'$, or the cold equilibrium 
distribution majorizes the hot equilibrium distribuiton
in the Otto cycle. 

The relation  $P\prec P'$ implies $M_i \geq 0$ in 
Eq. (\ref{qhc}).
So, we conclude that 
$Q_h \geq 0$ along with $Q_c \leq 0$, i.e. heat is absorbed (released)
by the system at the hot (cold) reservoir---the necessary conditions for the operation of an engine.
Then, from the first law of thermodynamics,  
the net work extracted per cycle, 
$\mathcal{W} = Q_h + Q_c \geq 0$, is given by  
\begin{equation}
    \mathcal{W} = \sum_{i=1}^{n-1}(\omega_i -\omega_i') M_i.
\label{work}
\end{equation}
Thus, given that $M_i \geq 0$, 
in order to extract net work (${\cal W} > 0$), 
at least one gap must shrink ($\omega_j >\omega_j'$) 
during the first adiabatic stroke. 
Clearly, this is a necessary, but not a sufficient
condition in the presence of a majorization relation. 

The efficiency of the Otto engine is defined as
\begin{equation}
\eta = \frac{\mathcal{W}}{Q_h} = 1 + \frac{Q_c}{Q_h},
\label{eta}
\end{equation}
which is bounded by the Carnot efficiency, in
consistency with the second law. This only implies 
that the total entropy generated by the 
heat cycle is positive: $\Delta {\cal S}_{\rm tot}  = 
-Q_c/T_c -Q_h/T_h \ge 0$, which yields $\eta \leq 1-T_c/T_h$. 
Since the Otto cycle involves the isochoric steps which 
are irreversible, so the entropy generated is 
positive definite, implying that the efficiency
is strictly less than the Carnot bound.
However, as we show below, we can find a tighter 
upper bound for the Otto efficiency, which is solely dependent
on the energy spectrum, while being independent of the reservoir
temperatures. Such an upper bound has been derived 
earlier for specific cases in spin systems
\cite{Thomas2011, Venu2021}.

From Eqs. (\ref{qhc}) and (\ref{eta}), we can express the efficiency as 
\begin{equation}\label{etamm}
\eta =  1-\frac{\sum_{i=1}^{n-1}\omega_i'M_i}{\sum_{i=1}^{n-1}\omega_iM_i}.
\end{equation}

Again, due to $\omega_i, \omega_i' > 0$ and $M_i \geq 0$, we
have the following bounds:
\begin{equation}
  \underset{i}{\rm min} \left(\frac{\omega_i'}{\omega_i}\right) \leq   \frac{\sum_{i=1}^{n-1}\omega_i'M_i}{\sum_{i=1}^{n-1}\omega_iM_i} \leq 
  \underset{i}{\rm max} \left(\frac{\omega_i'}{\omega_i}\right).
\label{etalb}
\end{equation}
The lower bound in Eq. (\ref{etalb}) 
helps to bound the Otto efficiency from above, as  
\begin{equation}
\eta_{\rm ub}=1-\underset{i}{\rm min} \left(\frac{\omega_i'}{\omega_i}\right).   
\label{etaubw}
\end{equation}
The bound is explicitly calculated for 
a specific working medium 
in Sec. \ref{ccex}.
Physically, we expect the maximum value of $|Q_c|/{Q_h} =1$,
which implies a vanishing efficiency. This  
results from $T_{\rm max} = T_h$, or if there is some gap which 
stays constant $\omega_j'=\omega_j$. 
Then, the Otto efficiency is lower bounded by the zero value.
To summarize, for an arbitrary working medium, when 
the hot probability distribution $P$ is majorized by 
the cold probability distribution $P'$, 
i.e. $P\prec P'$, 
the upper bound of the Otto efficiency is given by Eq. (\ref{etaubw}). 
Moreover, due to Eq. (\ref{tmin}),  
the upper bound can also be expressed as  
\begin{equation}
    \eta_{\rm ub}=1- \frac{T_{\rm min}}{T_h}.
    \label{etaub}
\end{equation}
It is clear that for a meaningful upper bound, 
we must have $T_{\rm min} < T_h$, which
requires that 
at least one energy gap must shrink 
during the first adiabatic stroke.  
Further, we have seen that when this passive state is put in
contact with a (cold) reservoir at $T_c$, then heat will always flow from
the system to the reservoir, provided $T_{\rm min} > T_c$. 
This in turn implies
that $\eta_{\rm ub}$ respects the Carnot bound.

As may be expected, the upper bound 
can be inferred from the passive state resulting from the second 
adiabatic stroke (see Fig. \ref{QOC}a), by finding the corresponding virtual temperatures $\{T_i'\}$. In this case, we have the relation
 $T_{\rm max}' = T_c/\underset{i}{\rm min} (\omega_i'/\omega_i)$, yielding   
$\eta_{\rm ub} = 1-T_c/T_{\rm max}'$. 
Along with Eq. (\ref{etaub}), these two forms of the upper bound 
imply that  
the quasi-static Otto cycle satisfies the relation:   
$T_{\rm min} T_{\rm max}' = T_c T_h$. 
Interestingly, the upper bound can also be related to the classical Otto efficiency under special circumstances.  
Assuming the working medium as a classical ideal gas, the  
efficiency of the classical Otto cycle can be written in two
equivalent forms: 
 $\eta_{\rm cl} = 1- {T_2}/{T_h}
              = 1-{T_c}/{T_4}$ \cite{ZemanskyBook}, 
where $T_2$ ($T_4$) is the temperature
of the gas after the first (second)
adiabatic process (see Fig. \ref{QOC}b), which 
may be regarded analogous to the virtual temperature $T_{\rm min}$ ($T_{\rm max}'$)
for the respective passive state in the quantum Otto cycle. Note that in the classical case, the working medium
is in internal equilibrium during the adiabatic stages 
with a well-defined temperature at each instant,
whereas in the quantum cycle, it is, in general, a
non-equilibrium state even for the quasi-static cycle.

\section{Coupled-spins system: an example}
\label{ccex}
As an illustration for the upper bound of Otto efficiency,
we consider the working medium of the QOE to be a pair of spin-1/2 particles coupled through an anisotropic XY Heisenberg  interaction,
with the Hamiltonian (in units of $\hbar = 1$) \cite{Cacmak2019, Chayan2024}
\begin{equation}
H=B (\sigma_z^{(1)}\otimes I+I\otimes \sigma_z^{(2)})+J[(1+\gamma)\sigma_x^{(1)}\otimes\sigma_x^{(2)}+(1-\gamma)\sigma^{(1)}_y\otimes\sigma_y^{(2)}].
\end{equation}
The coupling strength along the $x$- and the 
$y$-direction, 
is respectively given by $J_x = J(1+\gamma)$ and $J_y = J(1-\gamma)$, where $0 \leq \gamma \leq 1$ is the anisotropy parameter.
The magnetic field strength $B$ along the $z$-direction is modulated during the adiabatic processes, while $J$ and $\gamma$ are held fixed throughout the cycle.

\begin{figure}[h!]
    \centering
    \includegraphics[width=0.8\linewidth]{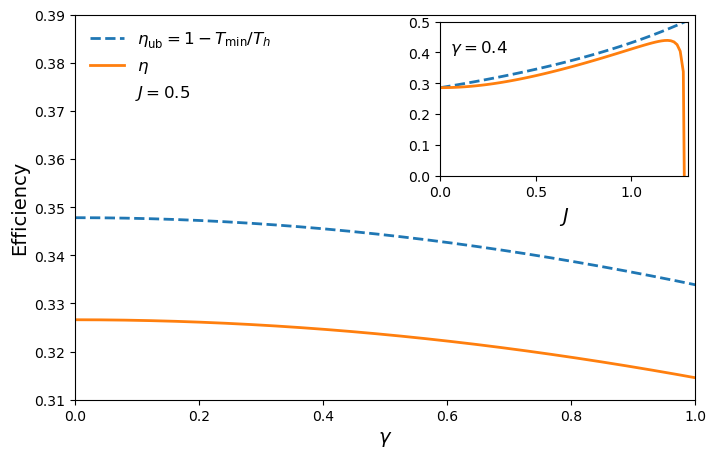}
    \caption{The quantum Otto efficiency ($\eta$) compared with its upper bound ($\eta_{\rm ub}$)  versus 
    the dimensionless anisotropy parameter $\gamma$
    in the XY Heisenberg interaction model,
    with $J=0.5$. 
    The inset shows $\eta$ and $\eta_{\rm ub}$ versus 
     $J$, with $\gamma=0.4$. 
     For both figures,  $B_1=2.8,B_2=2,T_h=1,T_c=0.5$.}
    \label{}
\end{figure}
The energy eigenvalues at the hot reservoir are given by 
$E_{1,4}=\mp2K_1$, $E_{2,3}=\mp 2J$  
and the corresponding eigenstates, in the computational basis, are 
\begin{align}
 \ket{\psi_{1,4}} &=\frac{1}{\sqrt{2}}\left(\frac{B_1\mp K_1}{\sqrt{K_{1}^{2}\mp B_1K_1}}\right)\ket{11}+\left(\frac{\gamma J}{\sqrt{K_{1}^{2}\mp B_1K_1}}\right)\ket{00}, \\
\ket{\psi_{2,3}} &=\frac{1}{\sqrt{2}}(\ket{01}\mp \ket{10}),
\end{align}
where we define $K_i=\sqrt{B_i^2+\gamma^2J^2}$, $i=1,2$.
The energy gaps at the hot reservoir are:
$\omega_1=2K_1-2J,~\omega_2=4J,~\omega_3=2K_1-2J$.

In the first adiabatic stroke, the external magnetic field 
decreases  from $B_1$ to $B_2$.
The virtual temperatures of 
adjacent levels for the 
resulting passive state are calculated as 
\begin{align}
 T_{1,3} &=T_h\frac{K_2-J}{K_1-J}, \quad T_2 =T_h,
\end{align}
which are ordered as $T_1=T_3<T_2$. So, the upper bound of the efficiency, 
$\eta_{\rm ub}=1-T_{\rm min}/T_h$, is 
explicitly given by 
\begin{equation}
\eta_{\rm ub} = 
\frac{\sqrt{B_1^2+\gamma^2J^2} - \sqrt{B_2^2+\gamma^2J^2} }{\sqrt{B_1^2+\gamma^2J^2}-J}.
  \label{etak12}
\end{equation}
Alternately, after the second adiabatic stroke, 
the virtual temperatures of adjacent levels, are given by
\begin{align}
T_{1,3}' = T_c\;\frac{K_1 - J}{K_2 - J}, \qquad
T_2' = T_c,    
\end{align}
which satisfy the ordering $T_1' = T_3' > T_2'$. In this case, the efficiency bound 
$\eta_{\rm ub} = 1 - {T_c}/{T_{\max}'}$ again yields Eq. (\ref{etak12}). 
Similarly, the present approach leads to the known upper bound in spins coupled through an isotropic Heisenberg interaction, as derived in Appendix \ref{XXXM} and reported earlier in Refs.
\cite{Thomas2011, Venu2021} 
without using the concept of virtual temperatures.   
The present approach establishes this bound in a simple way by relating
it naturally to the min-max values of the  virtual temperatures.

\section{Conclusions}
\label{Conc}
A passive quantum state can be characterized by
a set of positive virtual temperatures between adjacent energy levels of the system, which can be used to define a mean virtual temperature of the state. 
We have proposed this mean value    
as an indicator for comparison between passive states  
under isoenergetic and isoentropic transformations, 
analogous to the final temperatures 
of the equilibrium states obtained
under the corresponding classical transformations. Using majorization relations, we also show that 
the mean energy of the passive state
is bounded by the mean energies 
corresponding to the thermal states
at the min-max values of the 
virtual temperatures. These inequalities
then specify the  
temperature range allowing a unique direction of
 heat flow between the system
 and the environment put in mutual 
 thermal contact. Finally, we highlighted
 the role of virtual temperatures
 in setting an upper bound for
 the thermal efficiency of a quasi-static
 quantum  Otto engine. 
 Our results showcase virtual temperatures as a key operational quantity connecting passivity, majorization, and the performance of quantum thermal machines. We hope these insights will be valuable for appreciating  the similarities and differences between the classical and
 the quantum thermodynamic scenarios, understanding fundamental limits and guiding the design of future quantum thermal devices.
 Apart from
 the explicit example of a coupled
 qubit system, the present approach can 
 be extended to other working media
 as well as to other heat cycles
 including the refrigerator or 
 the heat pump modes. The possibility of 
 adapting this approach for finite-time
 cycles is an interesting future line
 of inquiry.

\section*{Acknowledgment}
S.S. acknowledges the Council for Scientific and Industrial Research (CSIR),  Government of
India for financial support via Award No. 09/947(0250)/2020-EMR-I.

\appendix

\section{Bounds for the mean energy $U(\rho^p)$}
\label{BoundME}
\par\noindent 
\textbf{Proof:} 
Consider a passive state $\rho^p$ of the system, with 
distribution $P = \{p_j\}_{j=1,...,n}$ for the occupation probabilities
of its (nondegenerate) energy levels. 
Let $T_i >0$, $i=1,...,n-1$, be the virtual temperature for every  
pair of adjacent levels separated by the gap 
$\omega_i = E_{i+1}-E_i$. 
We also consider a Gibbs distribution of the system,   denoted as
\(
Q(T_i) = (q_1(T_i), \ldots, q_n(T_i)), 
\)
for each temperature $T_i$. 
Then, we have 
\(
 {p_i}/{p_{i+1}} = e^{\omega_i/T_i} = {q_i(T_i)}/{q_{i+1}(T_i)}.
\)
Similarly,  ${q_i(T_{\min})}/{q_{i+1}(T_{\min})} = e^{\omega_i/T_{\rm min}}$.
Since $T_{\min} \le T_i$, we can write
\begin{equation}    
\frac{q_i(T_{\min})}{q_{i+1}(T_{\min})} \ge \frac{p_i}{p_{i+1}}, 
\qquad i = 1,\ldots,n-1. \tag{A1}
\end{equation}
The above set of inequalities can be rearranged as a chain of inequalities:
\begin{equation}
\frac{q_1(T_{\min})}{p_1} \ge \cdots \ge \frac{q_n(T_{\min})}{p_n}. \tag{A2}
\end{equation}
Now, consider the mediant \cite{mediant} of the ratios $q_j(T_{\min})/p_j$ above, where
$j=1,...,n$, given by
\(
{\cal M} = {\sum_{j=1}^n q_j(T_{\min})}/{\sum_{j=1}^n p_j}=1,
\)
where we use the normalization property of $P$ and $Q(T_{\min})$.
Then, according to the mediant inequality,   
\begin{equation}
\frac{q_1(T_{\min})}{p_1} \ge {\cal M}=1 \ge \frac{q_n(T_{\min})}{p_n}.
\tag{A3}
\end{equation}
So, we can assume that for some $j \le k < n$, we have $q_j(T_{\min}) \ge p_j$, while
$q_j(T_{\min}) \le p_j$ for $j \ge k+1$.
Now, let us define the quantity,
\(
\chi_m = \sum_{j=m+1}^{n} [  p_j -q_j(T_{\min}) ], 
\)
where $m=0,...,n-1$,  
and split it into two sets of terms as follows.
\begin{align}
\chi_m &= \sum_{j=m+1}^{k} \left[ p_j -q_j(T_{\min})  \right]
       + \sum_{j=k+1}^{n} \left[ p_j -q_j(T_{\min})  \right], 
       \nonumber \\
&\equiv A + B. \tag{A4}
\end{align}
It follows that $A <0$, since for $j \le k$, we have $p_j \leq q_j(T_{\min})$. Similarly, $B > 0$ also holds. Clearly, 
for $m = 0$, we have $\chi_0 = 0$, or $A = -B$.
In other words, $A$ and $B$ cancel each other for $m=0$. 
For $m \neq 0$, there are fewer negative terms 
that contribute to $A$ 
while $B$ remains the same. Note that  $k$ has a fixed value here.  So $B$ may outweigh $A$ for $m\neq 0$,
yielding $\chi_m \geq 0$. 
Since the given distributions are  ordered in the 
descending sense, the conditions
$\chi_m \ge 0$ define the majorization relation
$Q(T_{\min}) \succ P$,
indicating that $Q(T_{\min})$ is more ordered than $P$.
Thus, the occupation distribution $P$ of the given passive state 
$\rho^p$ is always majorized by the Gibbs distribution
corresponding to the minimum virtual temperature, $T_{\min}$.

Next, given the mean energies $U(\rho^p) = \sum_{i=1}^n E_i p_i$
and $U(\rho(T_{\min})) = \sum_{i=1}^n E_i  q_i(T_{\min})$, 
their difference  can be 
rewritten in terms of the energy gaps,  $\omega_m = 
E_{m+1} - E_m$, as  \cite{Sachin2025}:
\begin{align}
  U(\rho^p) - U(\rho(T_{\min}))
 & = \sum_{m=1}^{n-1} \omega_m \chi_m. \tag{A5}
\end{align}
As we have proved above $\chi_m \ge 0$,
and $\omega_m > 0$, so we obtain
\(
U(\rho_p) \ge U(\rho(T_{\min})).
\)
Thus, the existence of a majorization relation $Q(T_{\min}) \succ P$
directly implies that if the system in the initial state $\rho^p$ 
thermalizes with the environment at temperature $T_{\min}$, 
then the energy of the system will be lowered, or, in other 
words, heat will flow from the system to the environment.

Along similar lines, we can show that for $T_{\max} \ge T_i$, the distribution $P$
majorizes $Q(T_{\max})$, i.e. $P \succ Q(T_{\max})$,
implying 
\(
U(\rho(T_{\max})) \ge U(\rho_p),
\)
which also specifies the direction of the heat flow as 
being from the reservoir at temperature $T_{\max}$ to 
the system initially in the passive state $\rho^p$.
Upon combining the above two inequalities, we obtain Eq. (5).

\section{XXX model}
\label{XXXM}
We derive the upper bound ($\eta_{\rm ub}$) using a working medium of two spin-1/2 
particles coupled via XXX (isotropic) Heisenberg interaction.
The Hamiltonian is given by
$H=B(\sigma_{z}^{(1)}\otimes I+I\otimes \sigma_{z}^{(2)})+2J \vec{\sigma}^{(1)}.\vec{\sigma}^{(2)}$, 
where $B$ is the applied magnetic field and 
$0< J < B/4$ is the coupling strength.
The energy eigenstates are: $\ket{\psi_1}=\ket{11},\ket{\psi_2}=\frac{\ket{01}-\ket{10}}{\sqrt{2}},\ket{\psi_3}=\frac{\ket{01}+\ket{10}}{\sqrt{2}},\ket{\psi_4}=\ket{00}$, corresponding to eigenvalues: \(E_1=2J-2B,~E_2=-6J,~E_3=2J,~E_4=2J+2B\), respectively. So, the  energy gaps between adjacent levels are given by: $\omega_1=2B-8J,~\omega_2=8J,~\omega_3=2B$. 

In the first adiabatic stroke, the magnetic field 
is decreased as $B_1 \to B_2$. 
For the resulting passive state, 
the virtual temperatures 
$T_i = T_h (\omega_i'/\omega_i)$,
are given by 
\begin{align}
T_1 & = T_h\frac{B_2-4J}{B_1-4J}, \\
T_2 & = T_h, \\
T_3 & = T_h\frac{B_2}{B_1}.
\end{align}
Within the operation regime of the heat engine, 
the virtual temperatures are ordered as
$T_h=T_2>T_3>T_1 = T_{\rm min}$. 
Note that the condition, $T_{\rm min} < T_h$, 
is satisfied. 
Based on the previous discussion on the direction of heat 
flow, if $T_c < T_{\rm min}$,
then heat is always released to the cold reservoir
($Q_c <0$). Thus, the upper bound for 
the Otto efficiency 
for XXX model 
is given by: $\eta_{\rm ub} =1-T_{\rm min}/T_h = (B_1-B_2)/(B_1 - 4J)$  (see Ref. \cite{Thomas2011} for an alternate
derivation of this bound).


\begin{thebibliography}{99}
%
\bibitem{CasasVazquez2003}
J.~Casas-Vázquez and D.~Jou, 
{Temperature in non-equilibrium states: A review of open problems and current proposals},
\href{https://doi.org/10.1088/0034-4885/66/11/R03}
{Reports on Progress in Physics 66, 1937 (2003).} 

\bibitem{Hilbert2014}
S.~Hilbert, P.~H\"anggi, and J.~Dunkel, 
{Thermodynamic laws in isolated systems},
\href{https://doi.org/10.1103/PhysRevE.90.062116}
{Phys. Rev. E 90, 062116 (2014). }

\bibitem{Campisi2015}
M.~Campisi, 
{Construction of microcanonical entropy on thermodynamic pillars,}
\href{https://doi.org/10.1103/PhysRevE.91.052147}
{Phys. Rev. E 91, 052147 (2015).} 


\bibitem{Kliesch2014}
M.~Kliesch, C.~Gogolin, M.~J.~Kastoryano, A.~Riera, and J.~Eisert,
{Locality of temperature},
\href{https://doi.org/10.1103/PhysRevX.4.031019}
{Phys. Rev. X {4}, 031019 (2014).}

\bibitem{HernandezSantana2015}
S.~Hern\'{a}ndez-Santana, A.~Riera, K.~V.~Hovhannisyan, M.~Perarnau-Llobet, 
L.~Tagliacozzo, and A.~Ac\'{i}n,
{Locality of temperature in spin chains},
\href{https://doi.org/10.1088/1367-2630/17/8/085007}
{New J. Phys. {17}, 085007 (2015).}

\bibitem{Hartmann2004}
M.~Hartmann, G.~Mahler, and O.~Hess,
{Existence of temperature on the nanoscale},
\href{https://doi.org/10.1103/PhysRevLett.93.080402}
{Phys. Rev. Lett. {93}, 080402 (2004).}

\bibitem{Mahlerbook2008}
J.~Gemmer, M.~Michel, and G.~Mahler,
{\it Quantum Thermodynamics: Emergence of Thermodynamic Behavior Within Composite Quantum
Systems},
\href{https://doi.org/10.1007/978-3-540-70510-9}
{2nd ed., Lecture Notes in Physics, Springer: Berlin/Heidelberg, Germany (2009).}


\bibitem{Johal2009}
R.~S.~Johal,
{Quantum heat engines and nonequilibrium temperature},
\href{https://link.aps.org/doi/10.1103/PhysRevE.80.041119}
{Phys. Rev. E. {80}, 041119 (2009).}




\bibitem{Puglisi2017}
A.~Puglisi, A.~Sarracino, and A.~Vulpiani, 
{Temperature in and out of equilibrium: A review of concepts, tools and attempts},
\href{https://doi.org/10.1016/j.physrep.2017.09.001}
{Physics Reports 709–710, 60 (2017).} 


\bibitem{Allahverdyan2023}
A. E. Allahverdyan, S. G. Gevorkian, Yu. A. Dyakov, and Pao-Kuan Wang,
{Thermodynamic definition of mean temperature},
\href{https://link.aps.org/doi/10.1103/PhysRevE.108.044112}
{Phys. Rev. E 108, 044112 (2023).}


\bibitem{Hsiang2021}
J.-T.~Hsiang and B.-L.~Hu, 
{Nonequilibrium quantum free energy and effective temperature, generating functional, and influence action,}
\href{https://doi.org/10.1103/PhysRevD.103.065001}
{Phys. Rev. D 103, 065001 (2021).}




\bibitem{Sobolev2018}
S. L. Sobolev,
{Hyperbolic heat conduction, effective temperature, and third law for nonequilibrium systems with heat flux,}
\href{https://doi.org/10.1103/PhysRevE.97.022122}
{Phys. Rev. E 97, 022122 (2018).}


\bibitem{Ghonge2018}
S. Ghonge and D. C. Vural, 
{Temperature as a quantum observable,}
\href{https://iopscience.iop.org/article/10.1088/1742-5468/aacfb8}
{J. Stat. Mech. 073102 (2018).}

\bibitem{Skrzypczyk2015}
P.~Skrzypczyk, R.~Silva, and N.~Brunner,
{Passivity, complete passivity, and virtual temperatures},
\href{https://link.aps.org/doi/10.1103/PhysRevE.91.052133}
{Phys. Rev. E {91}, 052133 (2015).}


\bibitem{Pusz1978}
W.~Pusz and S.~L.~Woronowicz,
{Passive states and KMS states for general quantum systems}, \href{https://doi.org/10.1007/BF01614224}
{Communications in Mathematical Physics {58}, 273–290 (1978).}


\bibitem{Brunner2012}
N.~ Brunner, N.~Linden, S.~Popescu, and P.~ Skrzypczyk,
{Virtual qubits, virtual temperatures, and the foundations of thermodynamics},
\href{https://doi.org/10.1103/PhysRevE.85.051117}
{Phys. Rev. E 85, 051117 (2012).}



\bibitem{Silva2016}
R.~Silva, G.~Manzano, P.~Skrzypczyk, and N.~Brunner,
{Performance of autonomous quantum thermal machines: Hilbert space dimension as a thermodynamical resource},
\href{https://link.aps.org/doi/10.1103/PhysRevE.94.032120}
{Phys. Rev. E 94, 032120 (2016).}


\bibitem{Mitchison2019}
M.~T.~Mitchison,
{Quantum thermal absorption machines: refrigerators, engines and clocks},
\href{https://doi.org/10.1080/00107514.2019.1631555}
{Contemporary Physics 60(2), 164-187 (2019).}

\bibitem{Sparaciari2017}
C. Sparaciari, D. Jennings, and J. Oppenheim, Energetic instability of passive states in thermodynamics. \href{https://doi.org/10.1038/s41467-017-01505-4}
 {Nat Commun 8, 1895 (2017).}


\bibitem{Nikolaos2021}
N.~Koukoulekidis, R.~Alexander, T.~Hebdige, and D.~Jennings, 
{The geometry of passivity for quantum systems and a novel elementary derivation of the Gibbs state},
\href{https://doi.org/10.22331/q-2021-03-15-411}
{Quantum 5, 411 (2021).}

\bibitem{Brunner2023}
P.~Lipka-Bartosik, M.~Perarnau-Llobet, and N.~Brunner,
{Operational definition of the temperature of a quantum state},
\href{https://doi.org/10.1103/PhysRevLett.130.040401}
{Phys. Rev. Lett. {130}, 040401 (2023).}

\bibitem{JohalAJP2023} 
R. S. Johal, {The law of entropy increase for bodies in mutual thermal contact,}
\href{https://doi.org/10.1119/5.0124068}
{Am. J. Phys. 91 (1), 79–80 (2023).} 


\bibitem{NielsenBook} 
M. A. Nielsen and I. L. Chuang, {\it Quantum Computation and Quantum Information},
\href{https://michaelnielsen.org/qcqi/QINFO-book-nielsen-and-chuang-toc-and-chapter1-nov00.pdf}
{Cambridge university press (2000).} 

\bibitem{Sagawa2020}
T. Sagawa,
{\it Entropy, Divergence, and Majorization in Classical and Quantum Thermodynamics},  
\href{https://doi.org/10.1007/978-981-16-6644-5}
{Springer Briefs in Mathematical Physics;
Springer: Singapore (2020).}



\bibitem{Marshall2011}
A. W. Marshall, I. Olkin, and B. C. Arnold,
{\it Inequalities: Theory of Majorization and Its Applications},  
\href{https://doi.org/https://doi.org/10.1007/978-0-387-68276-1}
{Springer Series in Statistics; Springer:
New York, NY, USA (2011).}



\bibitem{Bhatia1996}
R. Bhatia,   
{\it Matrix Analysis},  
\href{https://doi.org/10.1007/978-1-4612-0653-8}
{Springer: New York, NY, USA (1996).}


\bibitem{Joe1990}
H. Joe, {Majorization and Divergence,} 
\href{https://doi.org/10.1016/0022-247X(90)90002-W}
{J. Math. Anal. Appl. 148, 287 (1990).}


\bibitem{Ruch1980}
E. Ruch, R. Schranner, and T. H. Seligman,
{Generalization of a theorem by Hardy, Littlewood, and Pólya,}
\href{https://doi.org/10.1016/0022-247X(80)90075-X}
{J. Math. Anal. Appl. 76, 222–229 (1980).}



\bibitem{Torun2023}
G. Torun, O. Pusuluk,  and Ö. E. Müstecaplıoğlu,
{A Compendious Review of Majorization-Based Resource Theories: Quantum Information and Quantum Thermodynamics,}
\href{http://dx.doi.org/10.55730/1300-0101.2744}
{Turkish Journal of Physics 47 (4), 141-182 (2023).} 

\bibitem{Nielsen2001}
M. A. Nielsen and  G. Vidal, 
{Majorization and the interconversion of bipartite states,}
\href{https://doi.org/10.26421/QIC1.1-5}
{Quantum Info. Comput. 1, 76–93 (2001).} 

\bibitem{Nielsen1999}
M. A. Nielsen,
{Conditions for a class of entanglement transformations,}
\href{https://doi.org/10.1103/PhysRevLett.83.436}
{Phys. Rev. Lett. 83, 436 (1999).} 

\bibitem{Horodecki2013}
M. Horodecki and J. Oppenheim, 
{Fundamental limitations for quantum and nanoscale thermodynamics,}
\href{https://doi.org/10.1038/ncomms3059}
{Nat. Commun. 4, 2059 (2013).}

\bibitem{Gour2018}
G. Gour, D. Jennings, F. Buscemi, R. Duan, and I. Marvian,
{ Quantum majorization and a complete set of entropic conditions for
quantum thermodynamics,}
\href{https://doi.org/10.1038/s41467-018-06261-7}
{Nat. Commun. 9, 5352 (2018).}

\bibitem{Uttam2021}
U Singh, S. Das, and N. J. Cerf, 
{Partial order on passive states and Hoffman majorization in quantum thermodynamics,}
\href{https://doi.org/10.1103/PhysRevResearch.3.033091}
{Phys. Rev.
Res. 3, 033091 (2021).}

\bibitem{Rethinasamy2020}
{S. Rethinasamy and M. M. Wilde,}
{Relative entropy and catalytic relative majorization,}
\href{https://doi.org/10.1103/PhysRevResearch.2.033455}
{Phys. Rev. Res. 2, 033455 (2020).}



\bibitem{Lostaglio2022}
{M. Lostaglio and K. Korzekwa,}
{Continuous thermomajorization and a complete set of laws for markovian thermal processes,}
\href{https://doi.org/10.1103/PhysRevA.106.012426}
{Phys. Rev. A 106, 012426 (2022).}


\bibitem{Alimuddin2019}
{M. Alimuddin, T. Guha, and P. Parashar,}
{Bound on ergotropic gap for bipartite separable states,}
\href{https://doi.org/10.1103/PhysRevA.99.052320}
{Phys. Rev. A 99, 052320 (2019).}


\bibitem{ADOJunior2022}
{A. D. O. Junior, J. Czartowski, K. Zyczkowski, and K. Korzekwa,}
{Geometric structure of thermal cones,}
\href{https://doi.org/10.1103/PhysRevE.106.064109}
{Phys. Rev. E 106, 064109 (2022).}



\bibitem{Sachin2023}
S.~Sonkar and R.~S.~Johal,
{Spin-based quantum Otto engines and majorization,
}
\href{https://doi.org/10.1103/PhysRevA.107.032220}
{Phys. Rev. A 107, 032220 (2023).}


\bibitem{Salvia4}
R.~Salvia and V.~Giovannetti, 
{Energy upper bound for structurally-stable N-passive states},
\href{https://doi.org/10.22331/q-2020-05-28-274}
{Quantum 4, 274 (2020). }



\bibitem{Llobet2015}
M.~Perarnau-Llobet, K.~V.~Hovhannisyan, M.~Huber, P.~Skrzypczyk, J. ~Tura, and A. Acín,
{Most energetic passive states},
\href{ https://doi.org/10.1103/PhysRevE.92.042147}
{Phys. Rev. E 92, 042147 (2015). }


\bibitem{Allahverdyan2004}
A.~E.~Allahverdyan, R.~Balian, and  Th.~M.~ Nieuwenhuizen,
{Maximal work extraction from finite quantum systems},
\href{ https://doi.org/10.1209/epl/i2004-10101-2}
{Europhys. Lett. 67 (4), 565–571 (2004). }



\bibitem{Ali2020}
M. Alimuddin, T. Guha, and P. Parashar,
{Independence of work and entropy for equal-energetic finite quantum systems: Passive-state energy as an entanglement quantifier},
\href{ https://doi.org/10.1103/PhysRevE.102.012145}
{Phys. Rev. E 102, 012145 (2020). }
\vspace{-0.5cm}
\bibitem{Chiribella2017}
G. Chiribella and Y. Yang,
{Optimal quantum operations at zero energy cost},
\href{ https://doi.org/10.1103/PhysRevA.96.022327}
{Phys. Rev. A 96, 022327 (2017). }


\bibitem{Narang2023}
V. Narang, R. Rai, and R. S. Johal,
Clausius' theorem and the second law in the process of isoenergetic thermalization,
\href{https://doi.org/10.1103/PhysRevE.110.054103}
{Phys. Rev. E {110}, 054103 (2023).}


\bibitem{Quan2007}
H. T. Quan, Yu-xi Liu, C. P. Sun, and F. Nori,
{Quantum thermodynamic cycles and quantum heat engines,}
\href{https://doi.org/10.1103/PhysRevE.76.031105}
{Phys. Rev. E 76, 031105 (2007).}

\bibitem{Fock1928}
M. Born and V. Fock,
{Beweis des Adiabatensatzes,
}
\href{https://doi.org/10.1007/BF01343193}
{Z. Fur Phys. 51, 165-180 (1928).}

\bibitem{Sachin2025}
 S. Sonkar and R. S. Johal,
Operational constraints in quantum Otto engines: Energy-gap modulation and majorization,
\href{https://doi.org/10.3390/e27060625}{Entropy
 27(6), 625 (2025).}



\bibitem{Thomas2011}
G. Thomas and R.~S.~Johal,
{Coupled quantum Otto cycle,
}
\href{ https://doi.org/10.1103/PhysRevE.83.031135}
{Phys. Rev. E 83, 031135 (2011).}


   

\bibitem{Venu2021}
R.~S.~Johal and V. Mehta,
{Quantum heat engines with complex working media, complete Otto cycles and heuristics,
}
\href{ https://doi.org/10.3390/e23091149}
{Entropy 23(9), 1149 (2021).}

\bibitem{ZemanskyBook}
M. W. Zemansky and R. H. Dittman,
{\it Heat and Thermodynamics. An Intermediate
Textbook,}
{New York, McGraw-Hill, Ch 6, 7th ed. (1997).}




\bibitem{Cacmak2019}
B. \ifmmode \mbox{\c{C}}\else \c{C}\fi{}akmak,  and  \"O. E. M\"ustecapl\ifmmode \imath \else \i \fi{}o\ifmmode \breve{g}\else \u{g}\fi{}lu,
{Spin quantum heat engines with shortcuts to adiabaticity,}
\href{ https://doi.org/10.1103/PhysRevE.99.032108}
{Phys. Rev. E 99, 032108 (2019).}





\bibitem{Chayan2024}
C. Purkait, S. Chand, and A. Biswas,
{Anisotropy-assisted thermodynamic advantage of a local-spin quantum thermal machine,}
\href{  https://doi.org/10.1103/PhysRevE.109.044128}
{Phys. Rev. E 109, 044128 (2024).}




\bibitem{mediant}
For $n$ given fractions, ordered as  \(
\frac{a_1}{b_1} \;\le\; \frac{a_2}{b_2} \;\le\; \cdots \;\le\; \frac{a_n}{b_n}, {\rm with} ~a_i>0,~b_i > 0, 
\)
the mediant inequality satisfies
\[
\frac{a_1}{b_1}
\;\le\;
\frac{\sum_{i=1}^n a_i}{\sum_{i=1}^n b_i}
\;\le\;
\frac{a_n}{b_n}.
\]





\end{thebibliography}
\end{document}